# Effects of Random Birefringence in Multimode Fibers on Nonlinear Ultrashort Pulse Propagation

Chaoyang Geng, Hengyu Liu, Lixia Xi, Xiaoguang Zhang, and Xiaosheng Xiao

*Abstract*—Nonlinear pulse propagation in multimode fibers (MMFs) has attracted significant attention recently due to the rich spatiotemporal nonlinearities and promising applications. In practical scenarios, random birefringence in MMFs cannot be neglected, affecting the polarization-dependent nonlinear pulse propagation. This paper investigates the influence of random birefringence in MMFs on nonlinear ultrashort pulse propagation using a modified generalized multimode nonlinear Schrödinger equation. Two scenarios, spatial beam self-cleaning and multimode soliton propagation, are specifically examined. It is found that while random birefringence typically weakens nonlinearity in MMFs, certain nonlinear processes such as soliton self-frequency shift caused by intra-pulse Raman effect exhibit a complex relationship with random birefringence. Moreover, the study reveals that beam self-cleaning can endure random birefringence at high input peak powers. This research provides guidance for practical applications that utilize the nonlinear transmission of ultrashort pulses in MMFs.

## I. Introduction

Imperfect fabrication of optical fiber and external environmental factors (e.g., fiber twisting or bending in experimental setups) cause the transverse profile of fibers to deviate from the ideal circular symmetry, inducing the effect of random birefringence. This random birefringence significantly impacts polarization-dependent nonlinear pulse propagation in fibers. Nonlinear Schrödinger equation (NLSE) are usually employed to describe nonlinear pulse propagation in single-mode fibers (SMFs) without considering polarization effects. When polarization effects due to random birefringence are significant, a two-component coupled NLSE becomes necessary [1]. Initial investigations focused on the interaction between nonlinearity and polarization effects during soliton propagation in SMFs [2], [3], and this was later extended to optical fiber communication scenarios [4]-[6].

In recent years, multimode fibers (MMFs) have rekindled interest primarily due to their larger mode area and additional spatial degrees of freedom compared to SMFs. MMFs can support numerous eigenmodes (i.e., spatial modes), enabling them to transmit more information or withstand higher pulse energies than SMFs. Space-division multiplexing techniques utilizing kilometers of MMFs have been employed in optical fiber communications to enhance transmission capacity [7]. In addition, nonlinear pulse propagation in short segments of passive MMFs has been extensively studied, revealing unique dynamics such as spatiotemporal instability [8]-[10], spatial beam self-cleaning [11], [12], and ultra-broadband supercontinuum generation [13], [14]. Since 2017, laser cavities based on MMFs have achieved spatiotemporal mode-locking, promising higher energy ultrashort pulse generation compared to SMF cavities [15]-[17]. For instance, output pulses with single pulse energies up to 150 nJ and durations of 150 fs were achieved in MMF cavities [15]. Furthermore, MMF cavities with spatial beam self-cleaning were demonstrated to improve the beam quality [18], [19].

Theoretical models for nonlinear pulse propagation in MMFs have been developed. F. Poletti and P. Horak derived the generalized multimode NLSE (GMMLSE) to accurately describe mode-resolved transmission properties, incorporating high-order effects such as dispersion, Raman scattering, and self-steeping terms [20]. L. G. Wright et al. developed a massively parallel numerical solver for rapid numerical solution of GMMNLSE, analyzing typical cases of nonlinear pulse propagation in MMFs [21]. Some studies have investigated nonlinear pulse propagation in MMFs considering polarization effects, particularly in weak- and strong-coupling regimes for space-division multiplexing transmission systems [22], multimode soliton propagation under strong mode coupling regimes [23], and intermodal four-wave mixing considering random birefringence fluctuations [24]. These studies mainly focused on optical fiber communication scenarios with fiber lengths on the order of kilometers and pulse durations on the order of picoseconds, where nonlinearity is primarily governed by the optical Kerr effect.

Herein, we focuses on the scenario of ultrashort pulse propagation in short segments (~meters) of MMFs, which has emerged recently in nonlinear fiber optics [8]-[14] and spatiotemporal mode-locking in MMF lasers [15]-[19]. In such scenarios, higher-order nonlinear effects such as intra-pulse Raman effects become significant. Moreover, intriguing nonlinear phenomena like spatial beam self-cleaning are observed [18], [19], which are rarely encountered in mode-division multiplexing communications. However, investigations into nonlinear ultrashort pulse propagation accounting for random birefringence in MMFs remain

This work was supported by the National Natural Science Foundation of China (No. 62375024); *(Corresponding author: Xiaosheng Xiao)*

The authors are with the State Key Laboratory of Information Photonics and Optical Communications, School of Electronic Engineering, Beijing University of Posts and Telecommunications, Beijing 100876, China (e-mail: xsxiao@bupt.edu.cn).

insufficient.

Extending our preliminary work [25], [26], this paper investigates the influence of random birefringence in MMFs on nonlinear ultrashort pulse propagation using a modified GMMLSE. The remainder of this paper is organized as follows. Section II describes the model for simulating nonlinear pulse propagation in MMFs with random birefringence. Section III presents the simulation parameters, while Section IV analyzes the effects of random birefringence on multimode nonlinearity, focusing on multimode soliton propagation with Raman effects and spatial beam self-cleaning. Finally, Section V concludes the paper.

## II. MODEL BASED ON GMMNLSE

To model random birefringence in MMFs, the fiber is discretized into numerous short sections, as illustrated in Fig. 1. Each section accounts for birefringence along its length and considers random rotation of the state of polarization (SOP) only at its interfaces [24]. It is assumed that within each section, the birefringence axes and dispersion parameters of all modes remain constant, akin to a polarization-maintaining fiber. However, adjacent sections have different birefringence axes orientations. This assumption is reasonable if each fiber section is short enough, which can be met in numerical simulations.

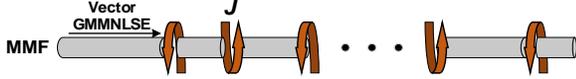

**Fig. 1.** Schematic diagram of the model.

For pulse propagation within each fiber section, a vector form of GMMNLSE is utilized. The GMMNLSE has been employed previously to describe nonlinear pulse dynamics in passive MMFs, incorporating high-order dispersion and nonlinear effects such as Raman and self-steepening [21]. Under the assumption of weak guidance, spatial modes in MMFs are often represented in terms of linearly polarized (LP) modes [27]. To include polarization effects here, each LP mode (e.g., $LP_{01}$, $LP_{11a}$, $LP_{11b}$…) is expressed using its Jones vector form, comprising $x$ and $y$-polarized components ($LP_{01}=LP_{01x}+LP_{01y}$, $LP_{11a}=LP_{11ax}+LP_{11ay}$, $LP_{11b}=LP_{11bx}+LP_{11by}$…), following similar treatments in previous studies [22-24]. This representation effectively doubles the number of modes, from $N$ modes in the scalar model to $2N$ polarized components.

By decomposing each scalar mode into two polarized modes along the birefringence axis of each fiber section, a slightly modified GMMNLSE is employed to describe polarization-dependent nonlinear pulse propagation within MMF sections:

$$\partial_z A_p(z,t) = i\delta\beta_0^{(p)} A_p - \delta\beta_1^{(p)} \partial_t A_p + \sum_{k=2}^{K} i^{k+1} \frac{\beta_k^{(p)}}{k!} \partial_t^k A_p$$
$$+ i\frac{n_2\omega_0}{c}\left(1+\frac{i}{\omega_0}\partial_t\right)\sum_{l,m,n}^{2N}\left[(1-f_R)S_{plmn}^K A_l A_m A_n^*\right. \quad (1)$$
$$\left.+ f_R S_{plmn}^R A_l \int_{-\infty}^{t} d\tau h_R(\tau) A_m(z,t-\tau) A_n^*(z,t-\tau)\right].$$

where $A_p(z,t)$ ($p = 1, 2…2N$) denotes the slowly varying temporal envelope of the light field for the $p$-th polarized mode, $\beta_k^{(p)}$ represents the $k$th-order dispersion for the $p$-th mode, $\delta\beta_{0,1}^{(p)} = \beta_{0,1}^{(p)} - \Re[\beta_{0,1}^{(0)}]$, $\Re$ denotes the real part only, $f_R$ is the Raman contribution to the Kerr effect (typically set to 0.18 for fused silica), $h_R$ is the Raman response of the fiber, and $n_2$ is the nonlinear index of refraction. The subscript $p$, $l$, $m$, and $n$ denote the polarized mode numbers: odd numbers represent $x$-polarized modes and even numbers represent $y$-polarized modes. $S_{plmn}^K$ and $S_{plmn}^R$ are the nonlinear coupling coefficients for the Kerr and Raman effects, respectively, and the simplification of them can be found in Appendix A.

For the optical field transmission between adjacent fiber sections, a projection matrix derived from Ref. [24] is used to couple polarized modes from the former section to the latter. Appendix B provides details of this matrix, which describes mode coupling across the entire mode family.

In numerical simulations, two characteristic lengths, $L_B$ (beat length) and $L_C$ (correlation length), are employed to characterize random birefringence in MMFs [24]. $L_B$ describes the birefringence within each fiber section, defined as the difference in propagation constants and group velocities between two perpendicular polarization modes (e.g., mode 1 and 2):

$$L_B = 2\pi / \left(\beta_{0,1}^{\text{mode 1}} - \beta_{0,1}^{\text{mode 2}}\right). \quad (2)$$

$L_C$, on the other hand, represents the correlation length that characterizes the degree of random SOP rotation, as defined in Appendix B. $L_C$ measures the typical distance over which random rotations of SOP start to become uncorrelated, with longer $L_C$ indicating less pronounced SOP rotation.

## III. SIMULATION PARAMETERS

Based on the above model, we will investigate the nonlinear ultrashort pulse propagation in single-core, graded-index MMFs. The refractive index profile of these MMFs is depicted in Fig. 2, and can be described by the equation:

$$n^2(r) = \begin{cases} n_0^2\left[1-2\Delta\left(\frac{r}{R}\right)^{\rho}\right], & r \le R. \\ n_1^2, & r > R. \end{cases} \quad (3)$$

where $\Delta = (n_0-n_1)/n_0$ represents the relative refractive index difference, $n_0$ and $n_1$ are the refractive indices of the core center and the cladding, respectively, and $\rho$ is the shape parameter that describes the rate of refractive index decrease. The parameters of the MMFs used in our simulations are detailed in Table I. The nonlinear refractive index $n_2$ is $2.3 \times 10^{-20}$ m$^2$W$^{-1}$. All mode profiles, dispersion parameters, and nonlinear coupling coefficients ($S_{plmn}^K$ and $S_{plmn}^R$) are calculated from the fiber specifications, as our and others' previous work [17], [21]. Up to fourth-order dispersions are considered in the simulations. Further details of simulation parameters are provided in the supplementary material.

The input pulses used in all examples are Gaussian-shaped. A massively parallel algorithm [21] is adopted to solve the vector GMMNLSE, Eq. (1). Note that, though $x$ and $y$-polarized components of LP modes are used in the numerical simulation, the *x*- and *y*-polarized components are combined to form each LP mode when we present the simulation results in the following figures.

TABLE I Simulation Parameters [a]

| | | Example. 1 | Example. 2 | Example. 3 |
|---|---|---|---|---|
| **Fiber parameters** | Length (m) | 0.5 | 15 | 0.5 |
| | Core diameter (μm) | 50 | 50 | 62.5 |
| | $n_1$ [b] | 1.4506 | 1.4449 | 1.4505 |
| | $n_0-n_1$ | 0.015 | 0.0137 | 0.0198 |
| | $\rho$ | 2.00 | 2.08 | 1.96 |
| | beat length, $L_B$ (m) | | in the order of 10 | |
| **Parameters of incident pulse** | Pulse duration (fs) | 150 | 50 | 60 |
| | Center wavelength (nm) | 1060 | 1550 | 1030 |
| | Total energy (nJ) | 50 | 6 | 440 |
| | Modes considered | 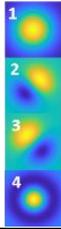 | 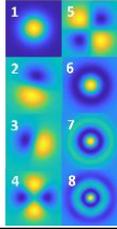 | 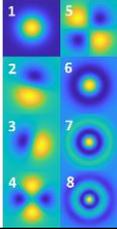 |

a. More details of the simulation parameters (dispersion parameters, nonlinear coupling coefficients, and beat length) are provided in the supplementary material.
b. The refractive index of the cladding at the center wavelength.

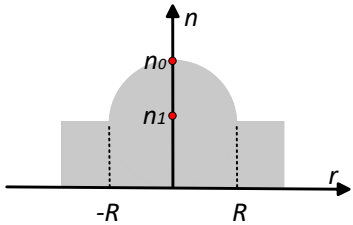

Fig. 2. Refractive index profile of the graded-index MMF, where $n_0$ and $n_1$ represent the refractive index of the core center and cladding, respectively.

## IV. EFFECTS OF RANDOM BIREFRINGENCE

Based on the model and simulation settings presented in sections II and III, respectively, we numerically investigate the effects of random birefringence in MMFs on nonlinearities through those three examples of Table I.

### A. Example 1: A Simple Case

Firstly, we apply the model to a simple example, launching a 150-fs multimode pulse into the MMF. For the input pulse, as depicted in Fig. 3(a,b), a total energy of 50 nJ is initially distributed equally across the first 4 modes (LP$_{01}$, LP$_{11a}$, LP$_{11b}$, LP$_{02}$) of the MMF. This simplified initial input condition allows for easier analysis, although more complex and realistic input conditions yield similar conclusions. Considering the characteristic lengths in common fibers, the beat lengths $L_B$'s of all modes are set to the order of magnitude of 10 m and the correlation length $L_C$ = 16.6 m. The mode-resolved temporal shape of input pulse and corresponding spectrum are shown in Figs. 3(a) and (b), respectively, and the simulation results of output without and with fiber random birefringence are illustrated in Figs. 3(c,d) and (e,f), respectively.

Comparing Figs. 3(e,f) with Figs. 3(c,d), the impact of random birefringence on nonlinear propagation is clearly evident. The pulse shapes in both temporal and spectral domains change significantly due to the introduced random birefringence. When random birefringence is present in the fiber, the spectral broadening of the pulse diminishes noticeably, primarily because nonlinear effects depend on the relative SOP of the light field. For instance, in the case of Kerr effect in the fiber, the components of the third-order susceptibility $\chi^{(3)}$ obey $\chi^{(3)}_{xxxx} = \chi^{(3)}_{xxyy} + \chi^{(3)}_{xyxy} + \chi^{(3)}_{xyyx}$, indicating that when interacting, the strongest nonlinear effects occur for light fields in the same linear polarization state. Additionally, birefringence causes the pulse components propagating along the fast and slow axes to separate in the time domain, leading to pulse broadening and reduced peak power. Therefore, under typical circumstances, nonlinearity in MMFs is generally weakened by random birefringence.

However, the interplay between nonlinear and linear effects in the fiber is complex. To further explore the interesting phenomena resulting from the effects of random birefringence, two typical nonlinear spatiotemporal dynamics described in Ref. [21] will be re-considered in the remaining part of this section, taking into account the random birefringence in MMFs.

effects, among which the intra-pulse Raman effect (i.e., soliton self-frequency shift) has a major contribution and the Kerr effect makes a minor contribution [21]. Modal walk-off is mitigated by cross-phase modulation among modes, aiding in the formation of a multimode soliton.

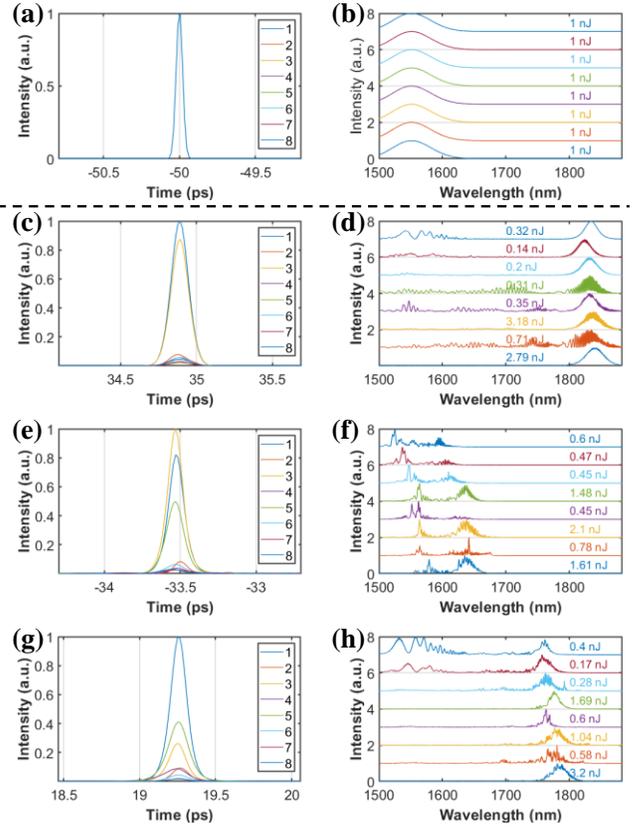

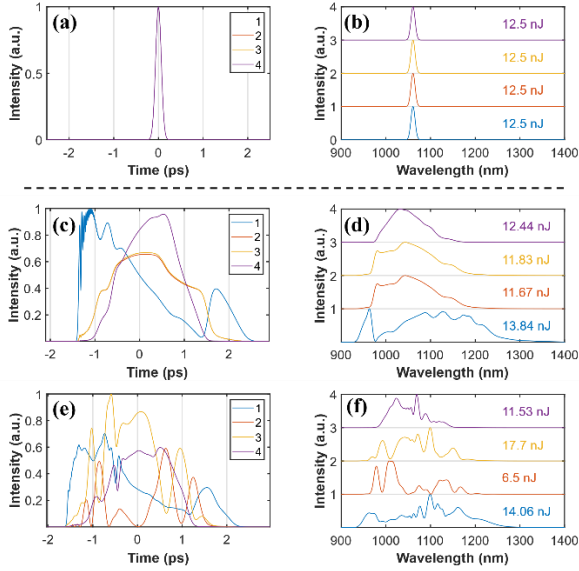

Fig. 3. Mode-resolved (a,b) input and (c-f) output pulses of the MMF. (c,d) without and (e,f) with random birefringence. (a,c,e) Temporal and (b,d,f) spectral domains. All four modes are overlapped in (a), and the 2nd and 3rd modes are overlapped in (c). Legends represent the number of modes. For better distinction of each mode in the spectral domain, the curve of the 2nd mode has been shifted upwards by 1 unit, and so forth, and the mode-resolved pulse energy is labelled.

Fig. 4. Mode-resolved (a,b) input and (c-h) output pulses of the MMF, (c,d) without, (e,f) with random birefringence ($L_C$ =166 m) and (g,h) with random birefringence ($L_C$ =16.6 m). (a,c,e,g) Temporal and (b,d,f,h) spectral domains. All eight modes are overlapped in (a). Legends represent the number of modes. For better distinction of each mode in the spectral domain, the curve of the 2nd mode has been shifted upwards by 1 unit, and so forth, and the mode-resolved pulse energy is labelled.

### B. Example 2: Multimode Soliton Propagation in MMF

Optical solitons are ultra-short pulses that maintain their shape and amplitude stability (for fundamental solitons) during propagation. Soliton formation is a balance between anomalous dispersion and nonlinear effects (typically due to self- and cross-phase modulations). In this simulation, the total energy is initially distributed uniformly across 8 modes, i.e., the 6 lowest-order modes and 2 other high-order modes with circular symmetry, the 15th and 28th modes (for convenience, they are numbered as modes 7 and 8 in Table I and Figs. 4). The center wavelength is set at 1550 nm, within the anomalous dispersion wavelength range. The polarization of the input pulse is along the slow axis of the first fiber section. Fig. 4(c) shows the multimode pulse in the temporal domain after propagating through a 15 m long MMF without random birefringence. Compared to the input shown in Fig. 4(a), the pulse center shifts from $t$ = -50 ps to $t \approx 35$ ps due to the combined effects of dispersion and frequency shift caused by nonlinear optical

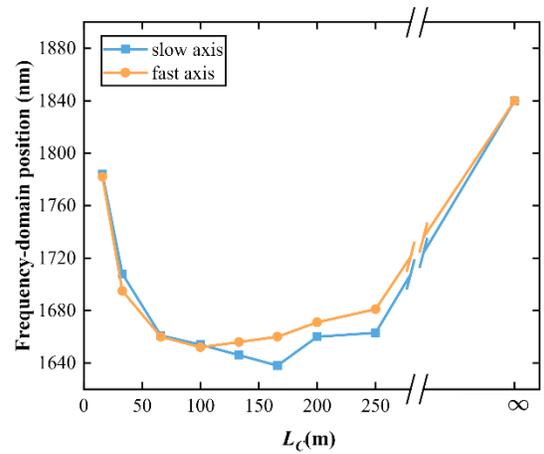

Fig. 5. Raman-induced wavelength shift varies with $L_C$.

Figs. 4(e,f) illustrate the outputs from the MMF with relatively weak random birefringence ($L_C$ =166 m). As shown in Fig. 4(f), the center wavelength of the pulse moves from 1550

nm to about 1640 nm. The Raman-induced frequency shift is dramatically smaller than the case without random birefringence. As a result, the temporal shift of the output pulse relative to the input significantly decreases in the MMF with random birefringence, as shown in the comparison of Fig. 4(e) (shift of ~16.5 ps) with 4(c) (~85 ps).

However, it is surprised to find that, as random birefringence becomes more intense (i.e., the correlation length $L_C$ decreases), the soliton self-frequency shift due to intra-pulse Raman effect increases. To show the impact of $L_C$ on Raman-induced frequency shift, the wavelength shift as a function of $L_C$ is depicted in Fig. 5. Two different input states are considered, i.e., the polarization of the incident light is along the fast and slow axes of the initial fiber section, respectively. It is found that regardless of the initial input SOP, the amount of Raman-induced frequency shift first decreases with decreasing $L_C$ (i.e., as random birefringence begins to appear), and then increases with further decrease in $L_C$ (i.e., more severe random birefringence). Our explanation for this is that birefringence causes the two polarization components propagating along the fast and slow axes to separate, leading to an increase in total pulse width and a decrease in peak power, thus reducing the Raman-induced soliton self-frequency shift. However, when $L_C$ is very small, i.e., the random SOP changes rapidly, the separation of the pulse components due to birefringence is weakened, resulting in minimal changes in peak power and making the Raman-induced frequency shift nearly identical to that with $L_C = \infty$ (i.e., without random birefringence), as the values at the leftmost ($L_C = 16.6$ m) and rightmost ($L_C = \infty$) in Fig. 5 are almost the same.

It is worth noting that the formation of multimode soliton remains, i.e., the modal walk-off of the soliton does not exhibit a significant change in the MMFs with random birefringence compared to the case of Fig. 4(c) without random birefringence. This demonstrates that multimode solitons possess a certain degree of resistance to the random birefringence in MMFs, which is consistent with the behavior of single-mode solitons [2].

*C. Example 3: Kerr Spatial Beam Self-cleaning Effect*

The spatial beam self-cleaning effect considering polarization has been recently investigated. Garnier et al. have provided insights into the mechanism underlying optical beam self-cleaning through the analysis of wave condensation in the presence of structural disorder (i.e., polarization random fluctuations) inherent to MMFs [28]. There, the temporal dispersion effects were neglected because nanosecond regime was considered. Nonlinear polarization dynamics have been experimentally associated with Kerr beam self-cleaning in graded-index MMFs [29]. The beam self-cleaning of different input SOPs has also been experimentally studied recently [30]. However, the impact of random birefringence in MMFs on nonlinear beam self-cleaning for ultrashort pulse inputs remains poorly understood. In this subsection, we analyze the effects of random birefringence in MMFs on spatial beam self-cleaning.

The total incident pulse energy is initially distributed uniformly across the 6 lowest-order modes, with noise present in the 7-10$^{th}$ modes. We assume the average difference between the refractive indices of the birefringence axes to be $\Delta n = 1.5 \cdot 10^{-7}$, which is a typical value for standard optical fibers. Other parameters are given in Table 1 and the supplementary material.

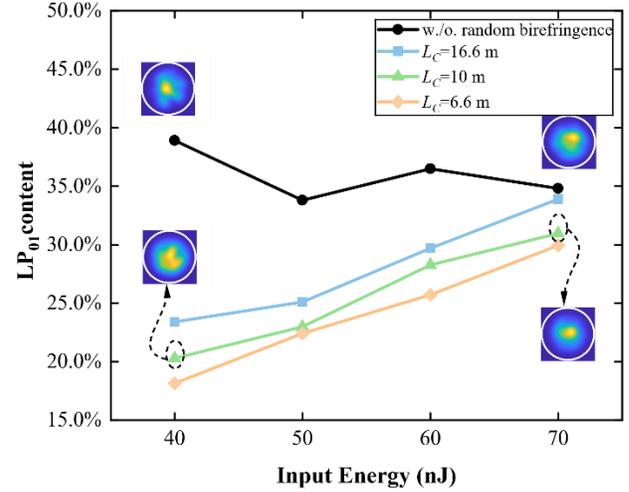

**Fig. 6.** Comparison of beam self-cleaning (representing by the proportion of LP$_{01}$ in the output beam) with and without random birefringence under different input peak power and different $L_C$. For each curve, different input pulse energies were investigated, with other parameters (including pulse width) remaining unchanged. Inset: the output beam profiles for pulse energy of 40 and 70 nJ, respectively, with or without random birefringence.

For the case of low pulse energy (40 nJ), we observe a significant degradation in the spatial beam self-cleaning due to the random birefringence in the MMF [26]. However, at high input energy levels, spatial beam self-cleaning can resist the effect of random birefringence in MMFs and yield improved output beam quality. Notably, within a certain threshold range, higher input power leads to a more pronounced resistance against random birefringence. This phenomenon is observed for MMFs with different parameters, and typical results are shown in Fig. 6. For example, when $L_C = 10$ m and the input pulse energy is increased from 40 to 70 nJ while keeping other parameters constant (i.e., increasing the peak power), there is a great improvement in the fundamental mode energy proportion, and the quality of the output beam is significantly enhanced, as shown in Fig. 6, where similar conclusions can also be drawn in other cases with different $L_C$. The results show that beam self-cleaning can resist the random birefringence in MMFs at high input peak power.

V. DISCUSSION AND CONCLUSION

The interplay between nonlinearity and polarization effects in SMFs or MMFs is a complex physical process, and this interaction is further complicated by the modal interplay in MMFs. Spatiotemporal nonlinearity of ultrashort pulse in MMFs has recently gained significant attention, with various interesting nonlinear dynamics being observed. However, investigations into the influence of random birefringence in MMFs on nonlinear effects are still insufficient. Apart from the simulation examples in Section IV, the modified GMMNLSE presented in Section II can also be utilized to explore the influences of the input SOP, polarization mode dispersion, etc., on multimode nonlinearity, and the nonlinear polarization dynamics along MMFs. Given the high degree of freedom in MMFs, we anticipate that more novel nonlinear phenomena within them will be discovered in the future.

In conclusion, this study has investigated the impacts of random birefringence in MMFs on the nonlinear propagation of ultrashort pulses, leveraging a modified GMMNLSE. Through a series of representative examples, we have examined how the random birefringence effect influences nonlinear spatiotemporal phenomena, contrasting scenarios with and without this effect. The findings indicate that, on the whole, random birefringence tends to mitigate nonlinear effects, aligning with observations made in SMFs. Nevertheless, the interplay between polarization and nonlinear effects in MMFs is particularly complex. The presence of random birefringence in common MMFs does not impede the formation of multimode solitons or beam self-cleaning when high peak power inputs are considered. For the case of multimode soliton, while moderate SOP rotation reduces the Raman-induced soliton self-frequency shift, substantial SOP rotation has little impact on the frequency shift.

## APPENDIX A

For each $x/y$ polarized mode, $S^K_{plmn}$ and $S^R_{plmn}$ can be calculated using the following equations [31]:

$$S^R_{plmn} = \frac{\int dxdy \left[\mathbf{F}^*_p \cdot \mathbf{F}_l\right]\left[\mathbf{F}_m \cdot \mathbf{F}^*_n\right]}{\left[\int dxdy |\mathbf{F}_p|^2 \int dxdy |\mathbf{F}_l|^2 \int dxdy |\mathbf{F}_m|^2 \int dxdy |\mathbf{F}_n|^2\right]^{1/2}}. \quad (A1)$$

$$S^K_{plmn} = \frac{2}{3} S^R_{plmn} + \frac{1}{3} \frac{\int dxdy \left[\mathbf{F}^*_p \cdot \mathbf{F}^*_n\right]\left[\mathbf{F}_m \cdot \mathbf{F}_l\right]}{\left[\int dxdy |\mathbf{F}_p|^2 \int dxdy |\mathbf{F}_l|^2 \int dxdy |\mathbf{F}_m|^2 \int dxdy |\mathbf{F}_n|^2\right]^{1/2}}. \quad (A2)$$

where $\mathbf{F}_p = \mathbf{e}_{x/y} F_p$ with the real-valued scalar mode function $F_p$ being the mode profile of the $p$-th spatial-mode, and $\mathbf{e}_{x/y}$ denotes the unit Jones vector of $x/y$ polarization. Clearly, the GMMNLSE is not a simple form, and an important consideration in determining the nonlinear terms of the equations is how to calculate the values of $S^K_{plmn}$ and $S^R_{plmn}$. Due to the orthogonality between the $x$- and $y$-polarized modes, the nonlinear coupling coefficients among different spatial modes may be 0, therefore greatly reducing computational complexity. Specifically, $S^R_{plmn}$ equals 0 except when $p$, $l$, $m$, and $n$ modes are in $xxxx/yyyy$ or $xxyy/yyxx$ polarization combinations. Calculating the values of $S^K_{plmn}$ is slightly more complex: when $p$, $l$, $m$ and $n$ are in $xyyx$ or $yxxy$ polarization combinations, $S^R_{plmn}$ equals 0, so $S^K_{plmn}$ can be calculated only by the second term on the right-hand side of equation (4); when $p$, $l$, $m$, and $n$ are in $xxxx/yyyy$ or $xxyy/yyxx$ polarization combinations, $S^K_{plmn}$ equals $S^R_{plmn}$ or $2S^R_{plmn}/3$, respectively; and in other combinations, $S^K_{plmn}$ equals 0.

## APPENDIX B

To calculate the mode coupling when light transmits through the interface of two adjacent fiber sections, we reference the fast axis of the $j$-th fiber section denoted as $\alpha_j(z)$, representing its angular orientation at position $z$ along the fiber. It is the electric field orientation of the mode that changes, while the shape of the mode remains unchanged. Subsequently, a $2N \times 2N$ projection matrix [24]:

$$\mathbf{P} = \begin{pmatrix} C & -S & 0 & 0 & 0 & 0 & 0 \\ S & C & 0 & 0 & 0 & 0 & 0 \\ \hline 0 & 0 & C^2 & -SC & -SC & S^2 & 0 \\ 0 & 0 & SC & C^2 & -S^2 & -SC & 0 \\ 0 & 0 & SC & -S^2 & C^2 & -SC & 0 \\ 0 & 0 & S^2 & SC & SC & C^2 & 0 \\ \hline 0 & 0 & 0 & 0 & 0 & 0 & \ddots \end{pmatrix} \quad (A3)$$

is applied to the polarized modes after every short distance propagation with lengths obeying a Gaussian distribution, where $C = \cos(\Delta\alpha)$ and $S = \sin(\Delta\alpha)$, with $\Delta\alpha = \alpha_{j+1} - \alpha_j$. Here, $N$ represents the number of vector modes. According to the matrix $\mathbf{P}$, coupling among different quasi-degenerate modes is induced by the random variation $\Delta\alpha$, and the linear coupling between the modes of different groups is neglected.

The correlation length $L_C$ is defined on the basis of the correlation function

$$C_\alpha(z) = \left|\int \alpha(z')\alpha(z'-z)dz'\right| / \int \alpha(z')^2 dz'. \quad (A4)$$

$L_C$ indicates the length beyond which the correlation function remains below 0.1.